\documentclass{PoS}
\title{Studying the Extreme Behaviour of 1ES~2344+51.4}
\ShortTitle{The Extreme Behaviour of 1ES~2344+51.4}

\author{\speaker{A.~Arbet-Engels} ${^1}$,
M.~Manganaro $^{2}$,
M.~Cerruti $^{3}$,
V.~Fallah Ramazani ${^4}$,
for the MAGIC\footnote{\texttt{https://magic.mpp.mpg.de/}. For collaboration list see PoS(ICRC2019)1177}~ Collaboration,
D.~Dorner $^{5}$ for the FACT\footnote{For collaboration list see PoS(ICRC2019)1177} ~Collaboration,
V.~Sliusar $^{6}$,
A.~V.~Filippenko$^{7,8}$,
T.~Hovatta$^{4}$,
V.~Larionov $^{9}$,
J.~A.~Acosta-Pulido $^{10,11}$,
C.~M.~Raiteri$^{12}$,
W.~Zheng$^{7}$\\
$^{1}$ ETH Zurich, CH-8093 Zurich, Switzerland, E-mail: \email{aaxel@phys.ethz.ch} \\ 
$^{2}$ University of Rijeka, Department of Physics, 51000 Rijeka, Croatia\\
$^{3}$ Universitat de Barcelona, E-08028 Barcelona, Spain\\
$^{4}$ Finnish Centre for Astronomy with ESO, FI-20014 University of Turku, Finland\\
$^{5}$ Universit\"at W\"urzburg, D-97074 W\"urzburg, Germany\\
$^{6}$ University of Geneva, Department of Astronomy, CH-1290 Versoix, Switzerland\\
$^{7}$ Department of Astronomy, University of California, Berkeley, CA 94720-3411, USA\\
$^{8}$ Miller Senior Fellow, Miller Institute for Basic Research in Science, University of California, Berkeley, CA  94720, USA\\
$^{9}$ Astronomical Institute of St. Petersburg State, 198504, St.Petersburg, Russia\\
$^{10}$ Instituto de Astrof\'isica de Canarias, Calle Via Lactea, s/n, E38205 La Laguna, Tenerife, Spain\\
$^{11}$ Departamento de Astrofisica, Universidad de La Laguna, E38205 La Laguna, Tenerife, Spain\\
$^{12}$ Osservatorio Astrofisico di Torino, I-10025 Pino Torinese, Italy 
}

\abstract{The BL Lac type object 1ES~2344+51.4 (redshift $z=0.044$) was one of the first sources to be included in the extreme high-peaked BL Lac (EHBL) family. EHBLs are characterised by a broadband spectral energy distribution (SED) featuring the synchrotron peak above $\sim 10^{17}$~Hz. From previous studies of 1ES~2344+51.4 in the very-high-energy (VHE, $>$100 GeV) gamma-ray range, its inverse Compton (IC) peak is expected around 200~GeV.
1ES~2344+51.4 was first detected in the VHE range by Whipple in 1995 during a very bright outburst showing around 60\% of the flux of the Crab Nebula above 350~GeV. In 1996, during another flare in the X-ray band, observations with Beppo-SAX revealed a large 0.1-10~keV flux variability on timescales of a few hours and an impressive frequency shift of the synchrotron peak to above $10^{18}$ Hz. Later on, this extreme behaviour of the source motivated several multiwavelength campaigns, during most of which the source appeared to be in a low state and showing no clear signs of ``extremeness''. In August 2016, FACT detected 1ES~2344+51.4 in a high state and triggered multiwavelength observations. The VHE observations show a flux level similar to the historical maximum of 1995. The combination of MAGIC, FACT, and Fermi-LAT spectra provides an unprecedented characterisation of the IC peak. It is the first time that simultaneous HE and VHE data are presented for this object during a flaring episode. We find an atypically hard spectrum in the VHE $\gamma$-rays as well as a hard X-ray spectrum, revealing a renewed extreme behaviour.}

\FullConference{36th International Cosmic Ray Conference -ICRC2019-\\
                July 24th - August 1st, 2019\\
                Madison, WI, U.S.A.}

\begin{document}
\section{Introduction}
Blazars constitute a prominent class of sources in the extragalactic very-high-energy (VHE, $>$100 GeV) sky\footnote{http://tevcat2.uchicago.edu/}. Belonging to the group of active galactic nuclei (AGN), they are characterized by a relativistic plasma jet pointing toward the observer. 
Blazars are commonly divided into two wide categories, flat spectrum radio quasar (FSRQ) and BL Lac type objects, according to the properties of their optical spectrum. FSRQs are identified by strong optical emission lines, while BL Lac type objects exhibit spectra with none or a very weak presence of such spectral features.\par
The spectral energy distribution (SED) of BL Lacs typically displays a two-hump structure. While the low-energy hump is generally attributed to synchrotron radiation emitted by highly relativistic electrons, the origin of the higher energy hump is still under debate. Most of the time, the latter component is successfully explained by synchrotron self-Compton (SSC) models, where the electron population responsible for the synchrotron radiation up-scatters the same synchrotron photons to GeV-TeV energies via the inverse-Compton process \cite{2007Ap&SS.309...95B}. However, other scenarios such as hadronic or lepto-hadronic models have been proposed \cite{2007Ap&SS.309...95B}. BL Lacs can further be divided into three subcategories depending on the location of the synchrotron peak $\nu_{synch, peak}$: low-energy peaked BL Lacs (LBL; $\nu_{synch, peak} < 10^{14}$ Hz ), intermediate-energy peaked BL Lacs (IBL; $10^{14}$ Hz $< \nu_{synch, peak} < 10^{15}$), and high-energy peaked BL Lacs (HBL; $10^{15}$ Hz $< \nu_{synch, peak} < 10^{17}$). During the past decade, however, X-ray observations have revealed that a small set of BL Lacs show a $\nu_{synch, peak}$ that is shifted to unusually high frequencies, above $10^{17}$ Hz. Based on this extremeness, \cite{Costamante2001} suggested the existence of an additional class of BL Lac objects, dubbed as extreme high-energy BL Lacs (EHBLs). Having a synchrotron peak located at higher energies, one expects the inverse-Compton bump to be also moved to higher energies, peaking at VHE. Consequently, EHBLs typically exhibit a hard spectrum with a photon index $\Gamma \lesssim 2$ both in the soft X-rays and in VHE.\par
These extreme conditions render the modeling of EHBL challenging. For instance, within an SSC scenario, the electron energy distribution (EED) must extend to the $\sim$TeV energies. Additionally, a hard TeV spectrum suggests a high minimum electron Lorentz factor for the electron population \cite{2011A&A...534A.130K}. Finally, the model parameters describing the SED often require a very low magnetization in the jet, suggesting that the energy budget in the emission region is far from equipartition between matter and magnetic field.\par
Recent observations suggest very different temporal behaviours among the EHBL family. Some of them, such as the classical EHBL 1ES~0229+200, seem to constantly display an extreme behaviour. On the other hand, several sources behave as EHBL only on a temporary basis and/or during flaring episodes (e.g., Mrk~501, see \cite{2018A&A...620A.181A}). 1ES~2344+51.4 belongs to the latter group. In this work, we report multiwavelength observations of 1ES~2344+51.4 during an intense flaring state that occurred in August 2016. The flare is characterized by hard VHE and X-ray spectra that we interpret as a renewal of extreme behaviour in these two energy bands.\par

\section{1ES~2344+51.4}
1ES~2344+51.4 is a nearby blazar located at a redshift of $z=0.044$  \cite{1996ApJS..104..251P}, and it is among the first extragalactic sources detected in the VHE band. Its discovery in the VHE $\gamma$-ray range was reported by the Whipple collaboration in 1995 \cite{Catanese_1998} during a flare, where the flux above 350 GeV reached $\sim$~0.6 Crab Units (C.U.)\footnote{For a given energy threshold, C.U. is defined as the integral flux of the Crab Nebula above the threshold energy.}, and the spectrum could be well described by a power law with an index of $2.54\pm0.17$ \cite{2005ApJ...634..947S}. Later on, \textit{BeppoSAX} observations performed during another high X-ray state in 1996 showed strong 0.1-10 keV flux variability on $\sim$hour timescale, together with impressive spectral variability \cite{2000MNRAS.317..743G}. A lower limit to the synchrotron peak frequency was set to $3\times10^{18}$ Hz, implying a frequency shift of the synchrotron peak by a factor $\sim$ 30 or more with respect to low states, which connotated this source for the first time as EHBL \cite{2000MNRAS.317..743G}. Interestingly, most recent multiwavelength campaigns probed the source mostly during quiescent states and not displaying the similar extremeness in the X-ray band seen during the 1996 flare, which suggests that extreme behaviours occur only during flaring states.

\section{Observations}
The First G-APD Cherenkov Telescope (FACT) is located at La Palma, Spain (altitude 2200~m) and measures photons at VHE using the imaging air Cherenkov technique \cite{Anderhub_2013, Biland_2014}. Thanks to the excellent performance stability of its silicon photomultiplier camera, FACT is an ideal monitoring instrument for bright TeV blazars. FACT continuously monitors 1ES~2344+51.4 and collected more than 1700 hr of observation time for this object after almost 8 years of operations \cite{Dornerproc}.\par
On MJD~57610 (August 10 2016\footnote{All dates herein are in UT.}), the FACT Quick Look Analysis \cite{Dorner:2015jka}, a low-latency on-site analysis, detected the source in an enhanced state. Based on this, the FACT collaboration issued an alert to the community. The flux of that night obtained from an off-site analysis was consistent with $\sim$ 0.5 C.U. above $\gtrsim$ 810~GeV. Follow-up observations were performed by several other instruments such as the Major Atmospheric Gamma Imaging Cherenkov (MAGIC) telescopes \cite{2016APh....72...76A}.\par
\begin{SCfigure}
    \centering
	\includegraphics[width=0.65\columnwidth]{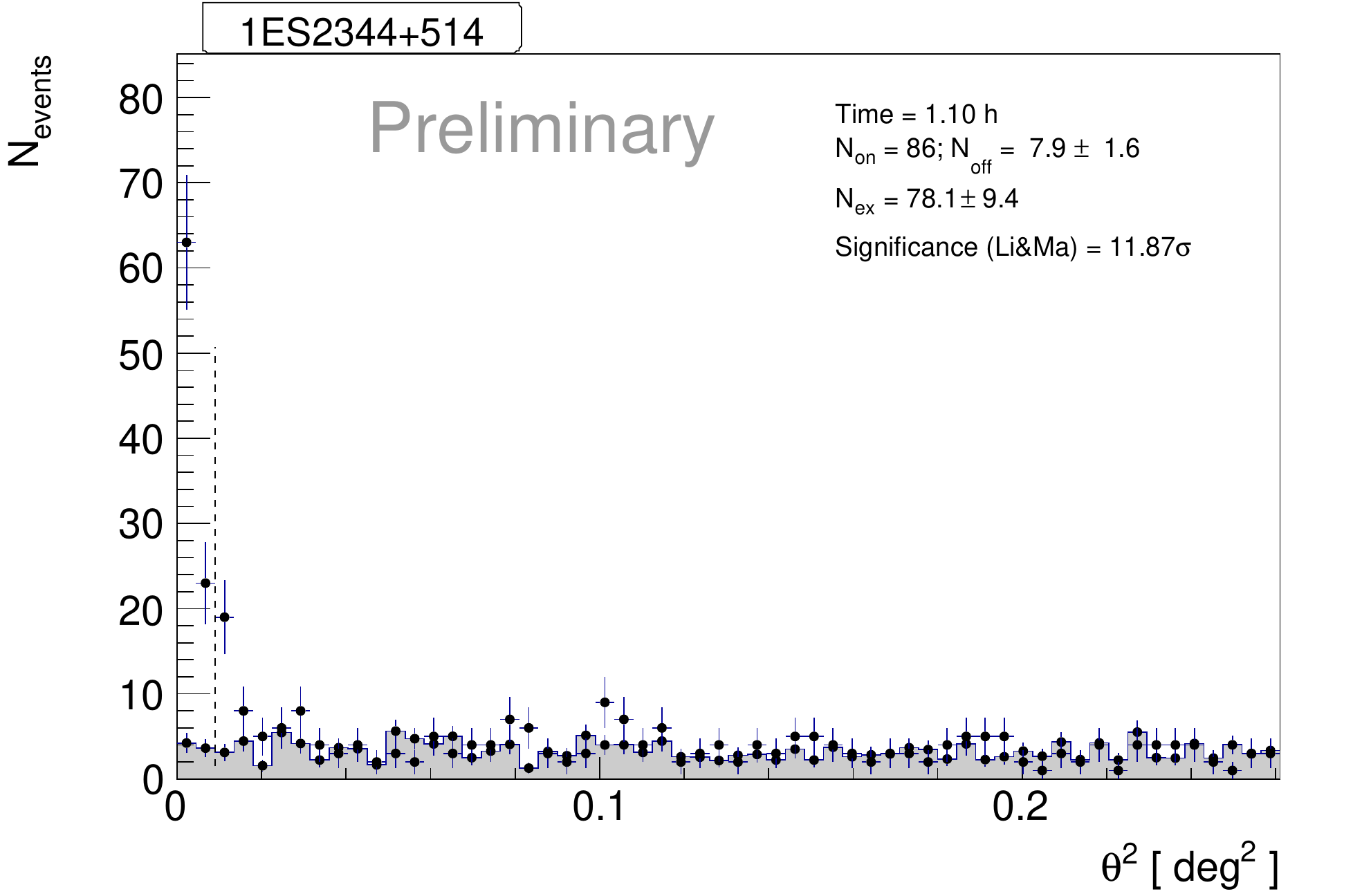}
    \caption{Distribution of the $\theta^{2}$ variable obtained from the overall MAGIC observation time. Dark points represent the $\theta^{2}$ values computed in the signal region, while the gray shaded area comes from the background region. The significance is calculated based on eq. 17 in \cite{1983ApJ...272..317L}, by considering all events on the left side of the dashed vertical line.}
    \label{fig:theta2}
\end{SCfigure}

MAGIC is a system of two 17-m imaging atmospheric Cherenkov telescopes located at La Palma, Spain, very close to FACT, and measures $\gamma$-rays above $E\gtrsim50$ GeV. MAGIC observed 1ES~2344+51.4 for two consecutive nights, MJD~57611 \& MJD~57612 (August 11 and August 12 2016), for a total of $\sim1.1$ hr, after quality cuts. A strong detection, significant at the $\sim12\,\sigma$ level, was obtained. In Figure~\ref{fig:theta2}, we show the distribution of the $\theta^2$ variable when combining the two observation nights. The $\theta^2$ variable denotes the angular distance (in [deg$^2$]) between the reconstructed direction and the expected source position of the gamma candidate events. The detection significance is calculated using eq. 17 in \cite{1983ApJ...272..317L}.\par
We complement VHE data with high-energy $\gamma$-ray (HE; $E>100$~MeV) observations provided by the Large Area Telescope (LAT) installed on the \textit{Fermi} satellite \cite{Atwood_2009}. For this work, we considered all \textit{Fermi}-LAT data taken between MJD~57567 \& MJD~57644 (June 28 \& September 13 2016). We then built the light curve between 300~MeV and 300~GeV and used a 7-day binning because of the faintness of the source in this band, typical of EHBLs. For this source, it is the first time that contemporaneous HE data are combined with VHE observations during a flare event, providing the best depiction of the IC peak so far. \par
X-ray observations were performed by the X-ray Telescope (XRT) onboard the \textit{Neil Gehrels Swift Observatory (Swift)} \cite{XRT_citation}. In total, five observations took place close to the MAGIC observations (MJD~57613, MJD~57620, MJD~57623, MJD~57626, and MJD~57629 -- August 13, 20, 23, 26,  and 29 2016). Additionally to XRT data, we analysed UV data provided by the Ultraviolet/Optical Telescope (UVOT), also onboard the \textit{Swift} satellite. All UVOT observations are strictly simultaneous to the XRT ones, and we considered the following three filters: UVW1, UVM2 \& UVW2.\par
1ES~2344+51.4 was also observed in the infrared (IR) and optical bands by several instruments: KVA, NOT (Tuorla blazar monitoring program\footnote{http://users.utu.fi/kani/1m/}), Stella, IAC80, AZT-8, and LX-200 (Whole Earth Blazar Telescope community\footnote{http://www.oato.inaf.it/blazars/webt/}) in the R-band. We also add observations from the KAIT telescope \cite{filippenko_li_treffers_modjaz_2001} that were performed without filter (i.e., in $clear$ band), which have an effective color close to the $R$ band. In the IR, we use data from the Telescopio Carlos S\'anchez (TCS) telescope in the $J$, $H$, and $K$-short filters. For the IR/optical range, we applied a host-galaxy correction following \cite{1999PASP..111.1223N}, since in these bands the host galaxy significantly contributes to the observed flux. \par
Finally, at the lowest energies, we use radio measurements at a frequency of 15 GHz, which were obtained by the OVRO 40-m telescope blazar monitoring program \cite{Richards_2011}. \par

\section{Results}
\begin{figure}
    \centering
    \makebox[\textwidth]{\includegraphics[width=1\columnwidth]{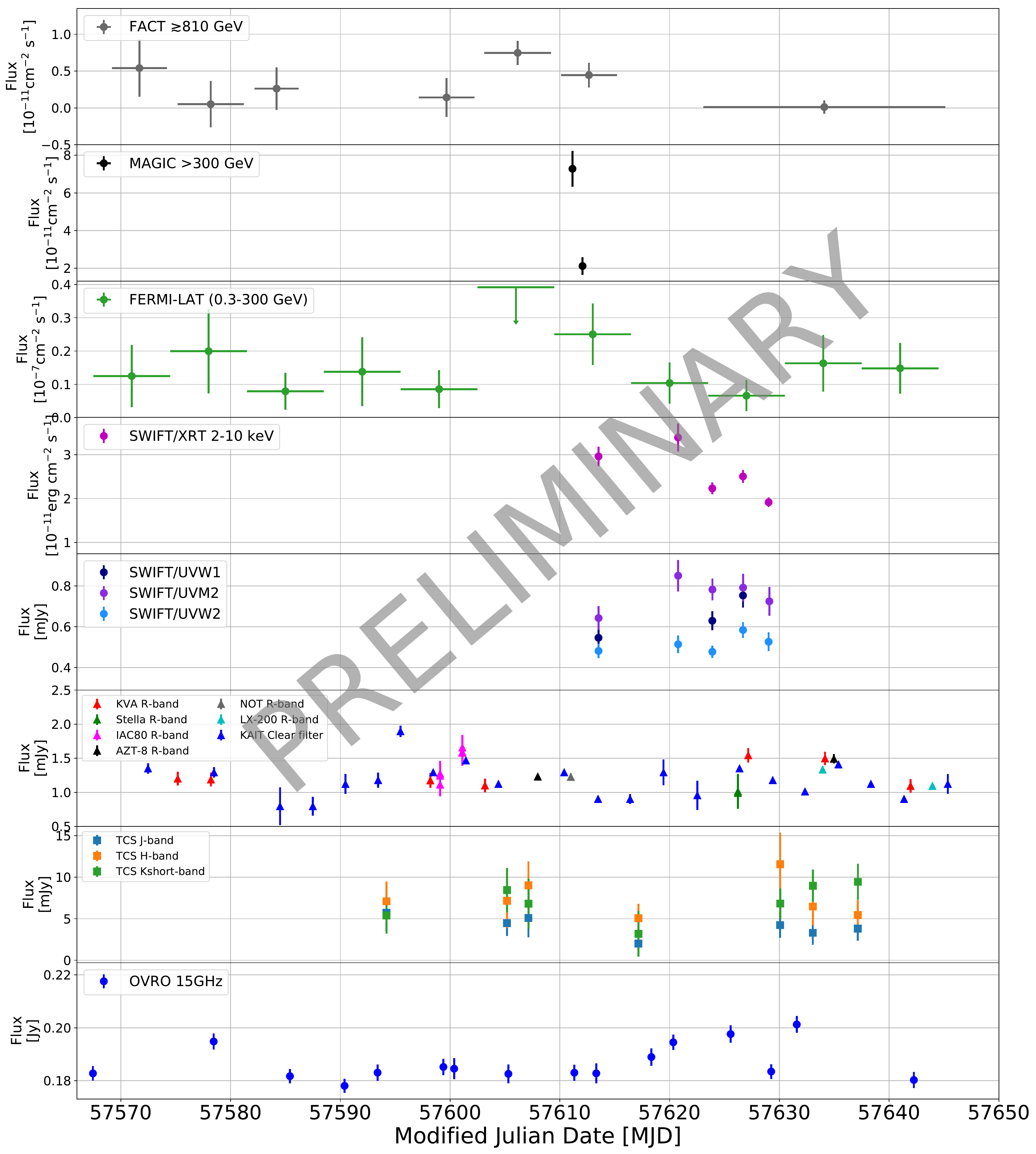}}
    \caption{Multiwavelength light curve of 1ES~2344+51.4 from MJD~57567 (28 June 2016) to MJD~57645 (14 September 2016). Light curves are obtained using (from top to bottom) FACT ($\gtrsim$810~GeV), MAGIC (>300~GeV), \textit{Fermi}-LAT (0.3-300~GeV), \textit{Swift}-XRT (2-10~keV), \textit{Swift}-UVOT (W1, M2, and W2 filters), KVA, Stella, IAC80, AZT-8, LX-200, NOT ($R$ band), KAIT (clear filter), TCS ($J$, $H$, $K$-short filters), and OVRO (15~GHz). In the \textit{Fermi}-LAT light curve, we quote an upper limit at 95\% C.L. for time bins with TS<3.}
    \label{fig:MWL_lc}
\end{figure}
We present the multiwavelength light curves from radio to VHE in Figure \ref{fig:MWL_lc}. All fluxes are computed in daily bins, except for the two monitoring instruments FACT \& \textit{Fermi}-LAT, where we use 7-day binning. Additionally, the last time bin of FACT is integrated over 1 month since the source faded and the measurements are consistent with no signal.\par
Between MJD~57603 and MJD~57615, the FACT light curve displays a weekly averaged flux of $\sim10^{-11}$cm$^{-2}$ s$^{-1}$ above $\gtrsim$~810~GeV, which is around $0.2$ C.U.\par
In the second panel from the top, we show the MAGIC light curve computed above $300$ GeV. During the first night, the measured flux is $F(>300$ GeV$)=(7.2\pm0.9)\times10^{−11}$cm$^{-2}$ s$^{-1}$, corresponding to $\sim0.55$ C.U. The flux is therefore similar to the historical maximum measured by Whipple in 1995 \cite{Catanese_1998}. On the second night, MJD~57612, the flux diminished by a factor of $\sim3.4$. Such a strong decrease brings clear evidence for a $\sim$daily scale variability at VHE energies. So far, no intranight variability has been observed at VHE in 1ES~2344+51.4.\par
The \textit{Fermi}-LAT light curve is shown in the third panel. As already mentioned, we present for the first time contemporaneous HE observations together with VHE data collected during a flaring episode. The source is relatively faint in this energy band, and was first detected after 5.5-months of \textit{Fermi}-LAT operations with a marginal test statistic TS $\approx$ 37 (\cite{2009ApJ...707.1310A}). Additionally, during previous multiwavelength campaigns no detection was claimed on timescales of $\sim$ months. On the contrary, over the considered period shown in the Figure \ref{fig:MWL_lc}, we report a strong detection with TS $\approx$ 110, revealing an overall enhanced HE state. Between MJD~57567.5 (28 June 2016) and
MJD~57644.5 (13 September 2016), the average flux was $F$(0.3-300 GeV)$=(1.2\pm0.2) \times 10^{-8}$ cm$^{-2}$ s$^{-1}$. The spectrum is best described with a power-law index of $\Gamma=1.9\pm0.1$. On the restricted 1-100 GeV band, the flux is $F$(1-100 GeV)$= (4.0 \pm 0.9) \times 10^{-9}$ cm$^{-2}$ s$^{-1}$, which is about two times higher than in the 3FGL catalog. Moreover, Figure \ref{fig:MWL_lc} shows an indication of a higher flux that is temporarily coincident with the flare seen in the VHE. The strong detection as seen by \textit{Fermi}-LAT provides an unprecedented constraint on the IC peak and will be exploited for the SED modelling.\par
The \textit{Swift}-XRT light curve in the 2-10 keV energy band is plotted in the fourth panel. The flux varies between $\sim 2 \times 10^{-11}$ erg cm$^{-2}$ s$^{-1}$ and $\sim3 \times 10^{-11}$ erg cm$^{-2}$ s$^{-1}$, and a decreasing trend is visible along the days. The VHE flare is therefore accompanied by elevated X-ray flux. Such a state remains, however, moderately high since the maximum flux that has been observed in this band is $\sim6$ erg cm$^{-2}$ s$^{-1}$ in December 2007 \cite{2011ApJ...738..169A}.\par

\begin{SCfigure}
	\centering
	\includegraphics[width=0.63\columnwidth]{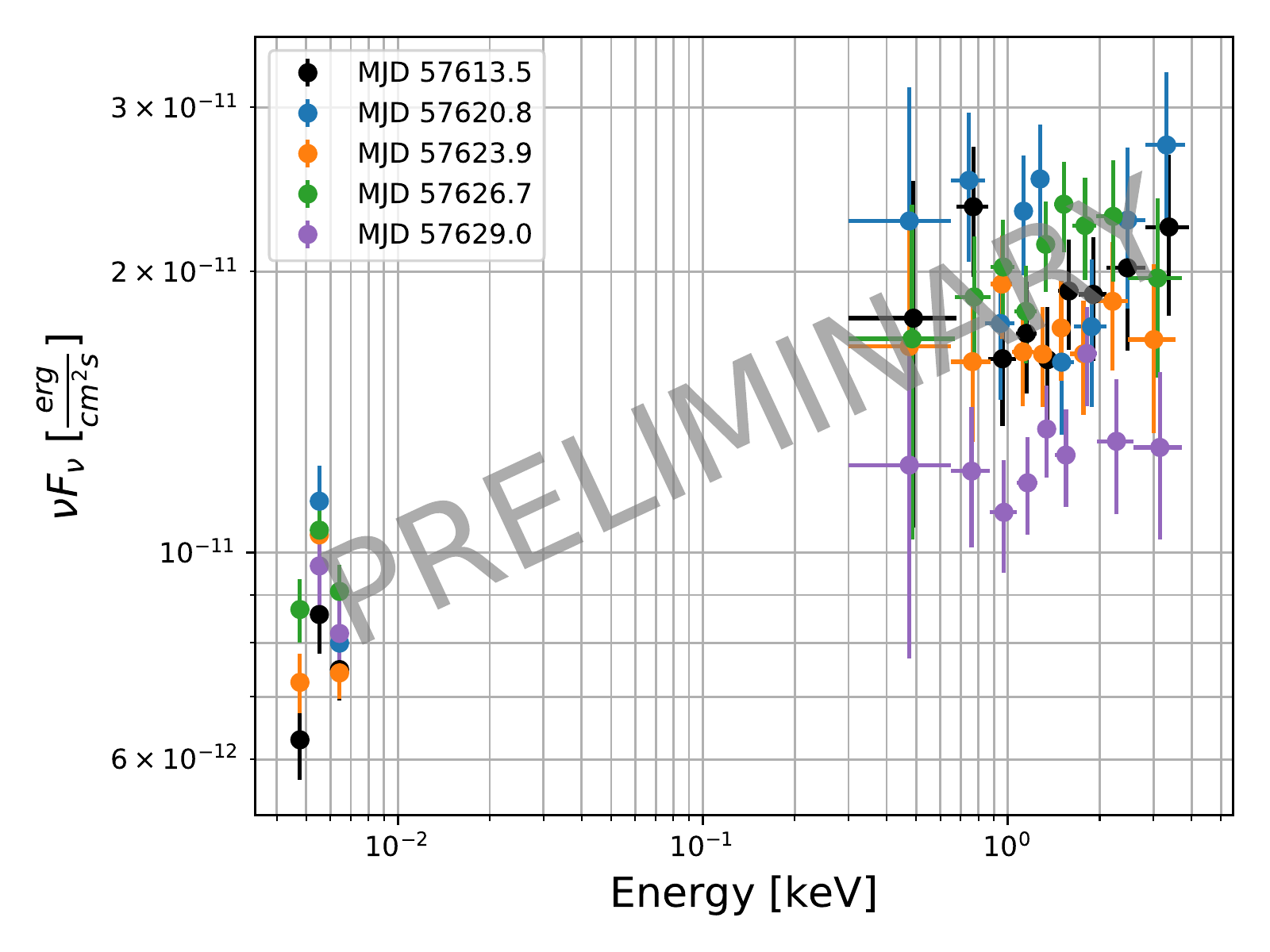}
    \caption{Strictly simultaneous combined UVOT-XRT SEDs from the five observations around the time of the MAGIC observations. All these observations are in agreement with an extreme synchrotron spectrum.}
    \label{fig:synch_sed}
\end{SCfigure}

Given the well-known X-ray spectral variability that 1ES~2344+51.4 exhibited in the past during high states, we studied the SED measured by \textit{Swift}-XRT. Furthermore, the location of the synchrotron peak is a crucial parameter, which defines the EHBL family. All \textit{Swift}-XRT spectra are well described with a power-law model. Additionally, we generally find hard spectra with a photon index $\Gamma_{XRT}\lesssim 2$. In Figure \ref{fig:synch_sed}, we plot the simultaneous combined UVOT-XRT SEDs from the five observations. Black data points represent the data taken on MJD~57613, which is one day after the MAGIC observations. For this night, the photon index is the hardest among the five observations and is equal to $\Gamma_{XRT}= 1.93 \pm 0.06$. Thus, Figure \ref{fig:synch_sed} and the hard photon index $\Gamma_{XRT}$ imply that UVOT and XRT data are describing the rising edge of the synchrotron bump, and we observe a shift of the synchrotron peak to $\gtrsim10^{18}$ Hz ($\gtrsim4$ keV) for this day. During low states, \cite{2000MNRAS.317..743G} and \cite{2013A&A...556A..67A} derived a peak below or equal to $\sim 10^{17}$ Hz. Consequently, we report a significant frequency shift of the low-energy hump by $\sim$1 order of magnitude. Such a behaviour is very similar to what was observed during the December 1996 high state X-ray \cite{2000MNRAS.317..743G}, and confirms the EHBL nature of the source occurring during flaring events. Within a SSC model, such a large shift of the synchrotron peak implies a displacement of the EED toward higher energies.\par 

\section{Conclusions}
We report multiwavelength observations of a strong flaring episode of 1ES~2344+51.4. The VHE flux was around $\sim0.55$ C.U. above $300$ GeV, comparable to the historical maximum observed by Whipple in 1995. A significant dimming is evident between the two MAGIC observations, revealing at VHE variability at $\sim$daily scale. The VHE flare was further accompanied with an enhanced X-ray state. Spectral analysis of the XRT data reveals a power-law spectrum with an index harder than 2, implying a frequency shift of the synchrotron peak to $\gtrsim10^{18}$ Hz. Our results confirm the EHBL behaviour already observed during flares. In an upcoming publication, a detailed analysis of the VHE spectrum will be presented. Moreover, taking advantage of the unprecedented constraint of the IC peak obtained when combining \textit{Fermi}-LAT data with VHE data, a modeling of the broadband emission will be presented and discussed. 

\footnotesize{
\section{Acknowledgments}
\noindent \textbf{MAGIC Collaboration:} {https://magic.mpp.mpg.de/acknowledgments\_ICRC2019/}. \textbf{FACT Collaboration:} {https://fact-project.org/collaboration/icrc2019\_acknowledgements.html}. The OVRO 40-m monitoring program is supported in part by NASA grants NNX08AW31G, NNX11A043G, and NNX14AQ89G, and NSF grants AST-0808050 and AST-1109911. This research has made use of data and/or software provided by the High Energy Astrophysics Science Archive Research Center (HEASARC), which is a service of the Astrophysics Science Division at NASA/GSFC and the High Energy Astrophysics Division of the Smithsonian Astrophysical Observatory. We acknowledge the use of public data from the Swift data archive. This article is based partly on observations made with the 1.5 TCS operated by the IAC in the Spanish Observatorio del Teide. This article is also based partly on data obtained with the STELLA robotic telescopes in Tenerife, an AIP facility jointly operated by AIP and IAC. A.V.F. and W.Z. are grateful for support from NASA grant NNX12AF12G, the Christopher R. Redlich Fund, the TABASGO Foundation, and the Miller Institute for Basic Research in Science (U.C. Berkeley). KAIT and its ongoing operation were made possible by donations from Sun Microsystems, Inc., the Hewlett-Packard Company, AutoScope Corporation, Lick Observatory, the US National Science Foundation, the University of California, the Sylvia and Jim Katzman Foundation, and the TABASGO Foundation. Research at Lick Observatory is partially supported by a generous gift from Google.

}

\bibliographystyle{JHEP}
\bibliography{bibliography_icrc_pos.bib}

\end{document}